\documentclass[a4paper,12pt]{extarticle}
\usepackage[a4paper,top=2cm,bottom=2cm,left=3cm,right=2cm]{geometry}
\usepackage{amssymb,graphicx}
\usepackage{epstopdf}
\usepackage{amsmath,amsfonts}
\usepackage{epsfig}
\usepackage{color}

\def\al{\alpha}
\def\eps{\epsilon}

\def\l{\left(}
\def\r{\right)}

\newcommand{\be}{\begin{equation}}
\newcommand{\ee}{\end{equation}}
\newcommand{\bea}{\begin{eqnarray}}
\newcommand{\eea}{\end{eqnarray}}
\newcommand{\bg}{\begin{gather}}
\newcommand{\eg}{\end{gather}}
\newcommand{\bseq}{\begin{subequations}}
\newcommand{\eseq}{\end{subequations}}

\def\half{\frac{1}{2}}

\newcommand{\tl}[1]{\tilde{#1}}
\newcommand{\tli}{\tilde{i}}%\imath}}
\newcommand{\tlj}{\tilde{j}}%\jmath}}
\newcommand{\ol}[1]{\overline{#1}}

\begin{document}
\begin{flushright}
%Prepint Number
\end{flushright}
\vspace{10pt}
\begin{center}
  {\LARGE \bf Interpreting multiple dualities conjectured from
    superconformal index identities} \\
\vspace{20pt}
%\medskip
A.~Khmelnitsky\\
\vspace{15pt}
\textit{
  Department of Physics, CERN - Theory Division,\\
  CH-1211 Geneva 23, Switzerland\\
  Institute for Nuclear Research of the Russian Academy of Sciences,\\
  60th October Anniversary Prospect, 7a, 117312 Moscow, Russia}\\
\end{center}
\vspace{5pt}

\begin{abstract}

We consider field theory side of new multiple Seiberg dualities
conjectured within superconformal index matching approach. We study
the case of $SU(2)$ supersymmetric QCD and find that the numerous
conjectured duals are different faces of handful of master theories.
These different faces are inequivalent to each other in a very
peculiar sense. Some master theories are fully known; we construct
superpotentials for others.  We confirm that all index identities
correspond to theories flowing to one and the same theory in the
infrared, thus supporting the conjecture of index matching for Seiberg
dual theories. However, none of the index identities considered in
this paper actually implies an entirely new, unknown duality.

\end{abstract}

%%%%%%%%%%%%%%%%%%%%%%%%%%%%%%%

\section{Introduction and summary}

New method of exploring Seiberg dualities has been suggested
recently. In Refs.~\cite{Romelsberger,Kinney}, the generalisation of
the Witten index for superconformal field theories was introduced, and
it was conjectured that the indices of theories related by Seiberg
duality should coincide~\cite{Romelsberger:2005eg}.  The coincidence
was checked for several known dual theories~\cite{Dolan,VSp:oct}.  The
index is a character of a relevant representation of certain subgroup
of superconformal group and counts ground states invariant under the
action of a particular supercharge. Thus, the conjecture seems natural
to hold for models flowing to the same infrared conformal fixed point.

Superconformal indices for gauge theories are given in terms of
elliptic hypergeometric integrals, and duality relations correspond to
their highly non-trivial transformation
properties~\cite{VSp:ell,Rains}.  Thus, the coincidence of indices is
new independent argument in favour of duality
conjecture. Comprehensive list of dualities and corresponding
relations for elliptic hypergeometric integrals, as well as
introduction to this recently emergent branch of special function
theory and its relation to Seiberg dualities can be found in
Ref.~\cite{VSp:nov}.

On the other hand, known transformation properties of elliptic
hypergeometric integrals lead to conjectures of new dualities between
supersymmetric gauge theories~\cite{VSp:oct,VSp:nov}. Although this
method provides only field content of conjectured duals, it must be
possible to construct complete field theories including their
superpotentials. A remarkable feature of the superconformal index
approach is that it suggests a multiplicity of duals to a single
``electric'' theory.  By making use of this approach, Spiridonov and
Vartanov~\cite{VSp:oct} (SV in what follows) have recently conjectured
71 dual descriptions for supersymmetric QCD with $N_c = 2$ colours and
$N_f = 4$ quark flavours whereas only three were known before
that. For the SQCD with $N_f = 3$ flavours, they have suggested 35
dual gauge theories. In the latter case, the low energy description in
terms of non-gauge theory was found in Ref.~\cite{Seiberg:exact}, but
no non-trivial dual gauge theory was known.

In this paper we study the phenomenon of multiple duals. We consider
duals conjectured by SV for $N_c=2$, $N_f=4,3$ SQCD and find that the
relationship between theories behind the superconformal index
identities can be called duality with reservations.  There are two
types of extra dualities.  The first one is inherent in electric
theories whose duals have enhanced accidental symmetries in the
infrared (for the discussion of enhanced symmetries in the Seiberg
duality context see, e.g., Ref.~\cite{Leigh}). Per se, these electric
theories have only a few ``master'' duals. Dualities proliferate once
some flavour current is coupled to external gauge field. Multiple dual
theories then have the same field content, but differ by the structure
of currents coupled to the external field.

Dualities of the second type are obtained from known ones by 
introducing a relevant operator into superpotential.  This
leads to new infrared behaviour of both electric and magnetic
theories obtained by integrating out some heavy fields or after
symmetry breaking. As the theories flow to the infrared, they can pass
through intermediate stages; some of these intermediate theories are
captured by the superconformal index approach.  These theories are
proper duals in the sense that they have the same infrared
description, but usually they are not considered as new non-trivial
independent dualities.

On the positive side, the superconformal index identities are
in remarkable correspondence with field theory dualities understood
in the above extended way. Field theories related by these identities 
do flow to the same infrared theory, and all index identities have
their field theory counterparts.
Thus, our
study can be considered as a check of the
conjecture that superconformal indices of Seiberg dual theories do
match and contain important group-theoretical information on the
structure of a theory.

The paper is organized as follows.  In Section 2 we summarize known
dualities of $N_c=2$, $N_f=4$ SQCD.  This theory has three ``master''
magnetic descriptions with different field contents and
superpotentials~\cite{Seiberg,Csaki,IP}. Although the original
electric theory has global symmetry group $SU(8)$, two of its duals
have lower symmetries. The latter symmetries get enhanced in the
infrared, where the full $SU(8)$ is restored~\cite{Leigh}. Hence, part
of $SU(8)$ global symmetry is accidental in magnetic descriptions.

Since part of $SU(8)$ is accidental in the two magnetic theories,
there is an ambiguity in identifying the operators of the electric
theory and their magnetic counterparts. We consider this point in
Section 3. The ambiguity becomes physical when the currents
corresponding to Cartan generators of $SU(8)$ are coupled to external
gauge fields. Numerous magnetic theories in external gauge fields
obtained in this way are inequivalent, and become identical in the
infrared only. This construction is in one-to-one correspondence with
SV counting based on superconformal index identities, hence giving the
interpretation of multiple SV dualities in $N_f=4$ theory.

In Section 4 we discuss $N_f=3$ theory as descendant of $N_f=4$.  We
will see explicitly that upon giving a mass to one electric quark
flavour, all four $N_f=4$ theories (electric and three magnetic) flow
to one and the same theory in the infrared. However, they do that
through different intermediate steps. These steps depend on the way
the quark mass term is introduced into the magnetic theory, so there
are several intermediate theories. Some intermediate theories are
precisely the duals suggested by SV via superconformal index approach.
In this way, and with account for the phenomenon described in Section
3, all SV duals are identified. Thus, we find that on the one hand,
the superconformal index technique does produce valid results, and on
the other hand, no entirely new non-trivial dualities are uncovered in
this way.

\section{$N_c=2,\, N_f=4$ SQCD and its duals}

Let us recall known properties of supersymmetric QCD with $SU(2)$
gauge group and $N_f =~4$ flavours of quarks, $Q^i$ in $\mathbf2$
representation and $\tl Q_{\tlj}$ in $\mathbf{\ol2}$ representation
($i, \tlj = 1,2,3,4$).  Its feature is that fundamental and
antifundamental representations of the gauge group are equivalent, so
``left'' quarks $Q$ and ``right'' quarks $\tl Q$ are combined into one
multiplet of $SU(8)$ flavour group.

The theory is believed to have non-trivial infrared fixed
point~\cite{Seiberg}. It has at least three Seiberg duals, i.e.,
theories which flow to the same fixed point in the infrared. All of
them have $SU(2)$ gauge group and at least $SU(4)_L\times SU(4)_R
\times U(1)_B$ global symmetry. They differ by field content, field
representations and superpotentials. The pattern of the dual theories
can be understood by considering their moduli spaces.

The moduli space is parametrized by all possible gauge invariants
(modulo classical relations implied by their definitions) giving
extremum to the superpotential.  In original electric theory, there is
no superpotential and moduli space is spanned by expectation values of
gauge invariants constructed from quarks. These are mesons $M^i_{\tlj}
= Q^i \cdot \tl Q_{\tlj}$, baryons $B^{ij} = Q^i \cdot Q^j$ and
antibaryons $\tl B_{\tli\tlj} = \tl Q_{\tli} \cdot \tl
Q_{\tlj}$. Because of enhanced flavour symmetry, they form together
antisymmetric tensor representation of $SU(8)$. Thus, all other
descriptions of this infrared fixed point must have the same moduli
space. In case of smaller flavour group, additional symmetry should
accidentally emerge in the infrared.

\begin{table}
\begin{tabular}{@{}cccc@{}}
\hline\hline
Field content & Quark gauge invariants & Superpotential & Moduli \\
\hline\hline
\rule{0cm}{1.1cm}
$\begin{array}{l r}
Q^i &  (\mathbf4,\mathbf1)^{+1} \\
\tl Q_{\tlj} & (\mathbf1,\mathbf4)^{-1} 
\end{array}$ &
$\begin{array}{l l}
B^{ij} \equiv Q^i \cdot Q^j & (\mathbf6,\mathbf1)^{+2}\\
\tl B_{\tli \tlj} \equiv \tl Q_{\tli} \cdot \tl
  Q_{\tlj} & (\mathbf1,\mathbf6)^{-2}\\
M^i_{\tlj} \equiv Q^i \cdot \tl Q_{\tlj} & (\mathbf4,\mathbf4)^0
\end{array}$ & &
$\l \begin{array}{r r}
B^{ij}& M^i_{\tlj}\\
-M^j_{\tli} & \tl B_{\tli \tlj}
\end{array}\r$ \\
\hline\hline\rule{0cm}{1.3cm}
$\begin{array}{c l}
q^i &  (\mathbf4,\mathbf1)^{-1} \\
\tl q_{\tlj} & (\mathbf1,\mathbf4)^{+1}\\
B^{ij} & (\mathbf6,\mathbf1)^{+2}\\
\tl B_{\tli \tlj} & (\mathbf1,\mathbf6)^{-2}\\
\end{array}$ &
$\begin{array}{l l}
C^{ij} \equiv q^i \cdot q^j & (\mathbf6,\mathbf1)^{-2}\\
\tl C_{\tli \tlj} \equiv \tl q_{\tli} \cdot \tl
  q_{\tlj} & (\mathbf1,\mathbf6)^{+2}\\
N^i_{\tlj} \equiv q^i \cdot \tl q_{\tlj} & (\mathbf4,\mathbf4)^0
\end{array}$ &
$\begin{array}{l}
\frac1{4\mu} \eps_{ijkl}B^{ij}\,q^k \cdot q^l +\\
\frac1{4\mu} \eps^{\tli \tlj \tl k \tl l}
\tl B_{\tli \tlj}\,\tl q_{\tl k} \cdot \tl
q_{\tl l}
\end{array}$&
$\l \begin{array}{r r}
B^{ij}& N^i_{\tlj}\\
-N^j_{\tli} & \tl B_{\tli \tlj}
\end{array}\r$ \\
\hline\rule{0cm}{1.1cm}
$\begin{array}{c l}
q_i &  (\mathbf{\ol4},\mathbf1)^{+1} \\
\tl q^{\tlj} & (\mathbf1,\mathbf{\bar4})^{-1}\\
M^i_{\tlj} & (\mathbf4,\mathbf4)^0
\end{array}$ &
$\begin{array}{l l}
C_{ij} \equiv q_i \cdot q_j & (\mathbf{\bar6},\mathbf1)^{+2}\\
\tl C^{\tli \tlj} \equiv \tl q^{\tli} \cdot \tl
  q^{\tlj} & (\mathbf1,\mathbf{\bar6})^{-2}\\
N_i^{\tlj} \equiv q_i \cdot \tl q^{\tlj} &
(\mathbf{\bar4},\mathbf{\bar4})^0 
\end{array}$ &
$\frac1\mu M^i_{\tlj}\,q_i \cdot \tl q^{\tl l} $&
$\l \begin{array}{r r}
\eps^{ijkl}C_{kl}& M^i_{\tlj}\\ -M^j_{\tli} &
\eps_{\tli \tlj \tl k \tl l}\tl C^{\tl k \tl l}
\end{array}\r$ \\
\hline\rule{0cm}{1.6cm}
$\begin{array}{c l}
q_i &  (\mathbf{\bar4},\mathbf1)^{-1} \\
\tl q^{\tlj} & (\mathbf1,\mathbf{\bar4})^{+1}\\
B^{ij} & (\mathbf6,\mathbf1)^{+2}\\
\tl B_{\tli \tlj} & (\mathbf1,\mathbf6)^{-2}\\
M^i_{\tlj} & (\mathbf4,\mathbf4)^0
\end{array}$ &
$\begin{array}{l l}
C_{ij} \equiv q_i \cdot q_j & (\mathbf{\bar6},\mathbf1)^{-2}\\
\tl C^{\tli \tlj} \equiv \tl q^{\tli} \cdot \tl
  q^{\tlj} & (\mathbf1,\mathbf{\bar6})^{+2}\\
N_i^{\tlj} \equiv q_i \cdot \tl q^{\tlj} &
(\mathbf{\bar4},\mathbf{\bar4})^0 
\end{array}$ &
$\begin{array}{c}
\frac1\mu M^i_{\tlj}\;q_i \cdot \tl q^{\tl l}+\\
\frac1{2\mu} B^{ij}\;q_i \cdot q_j +\\
\frac1{2\mu} \tl B_{\tli \tlj}\;
\tl q^{\tli} \cdot \tl q^{\tlj}
\end{array}$&
$\l \begin{array}{r r}
B^{ij}& M^i_{\tlj}\\
-M^j_{\tli} & \tl B_{\tli \tlj}
\end{array}\r$ \\
\hline
\end{tabular}
\caption{$N_c=2,\, N_f=4$ supersymmetric QCD and its duals.\label{Table}}
\end{table}

Electric SQCD and all three duals are described in Table~\ref{Table},
with the representations of the common global symmetry group
$SU(4)_L\times SU(4)_R \times U(1)_B$.  The original theory and the
third (last) dual possess $SU(8)$ flavour symmetry. For the first two
duals, it is impossible to arrange elementary fields in the $SU(8)$
multiplets, and flavour $SU(8)$ emerges accidentally in the infrared.

The tree duals are constructed by introducing to SQCD some of the
moduli of electric theory as elementary fields. Then one has to get
rid of similar composite moduli in order to restore the proper moduli
space structure. To this end, a superpotential is introduced.
Representations of magnetic quarks $q$, $\tl q$ are then completely
fixed by demanding the coincidence of moduli spaces. In order to match
elementary scalars to electric composites of canonical dimension 2, a
new energy scale $\mu$ is introduced.

The first dual description in our table was considered by Cs\'aki et
al.~\cite{Csaki} and contains baryons $B^{ij}$ and antibaryons $\tl
B_{\tli\tlj}$ of electric QCD as elementary fields. The presence of
these fields explicitly breaks $SU(8)$ symmetry down to $SU(4)_L\times
SU(4)_R \times U(1)_B$. There is the superpotential, which gives the
mass to composite magnetic baryons $C^{ij} \equiv q^i \cdot q^j$ and
antibaryons $\tl C_{\tli \tlj} \equiv \tl q_{\tli} \cdot \tl
q_{\tlj}$. This superpotential is crucial to match moduli spaces. The
moduli space of this dual theory is parametrized by $SU(8)$
antisymmetric tensor containing elementary baryons and composite
mesons\footnote{In electric theory, there are classical relations
between mesons and baryons. In $SU(8)$ language, they state that the
rank of the antisymmetric matrix parameterizing the moduli space does
not exceed 2. In magnetic theory, the rank of the moduli matrix is
constrained partly by the superpotential and partly by the fact that
baryon vev of rank greater than 2 leads to runaway vacuum.}
$N^i_{\tlj} \equiv q^i \cdot \tl q_{\tlj}$.  With this identification
of moduli spaces, one relates electric gauge invariants to magnetic
ones surviving in the infrared. Thus, one identifies baryons $B^{ij}$
and antibaryons $\tl B_{\tli\tlj}$ present in both theories, and
mesons $M^i_{\tlj}$ of electric theory with magnetic ones
$N^i_{\tlj}$. Composite baryons $C^{ij}$, $\tl C_{\tli\tlj}$ are
absent in the infrared because of the superpotential; they do not have
electric counterparts.

The generalisation of this duality to higher rank gauge groups has
been found recently by using superconformal index identities ($SU -
SP$ series of~\cite{VSp:nov}).

The second dual was considered in the original paper by
Seiberg~\cite{Seiberg} as part of the series of dual theories with
$SU(N_c)$ gauge groups. Electric mesons $M^i_{\tlj}$ are introduced as
elementary fields and, together with composite baryons $C_{ij} \equiv
q_i \cdot q_j$ and antibaryons $\tl C^{\tli \tlj} \equiv \tl q^{\tli}
\cdot \tl q^{\tlj}$, they parametrize the moduli space. Superpotential
takes care of now redundant composite mesons $N_i^{\tlj} \equiv q_i
\cdot \tl q^{\tlj}$. One identifies mesons $M^i_{\tlj}$ of electric
and dual theories, and electric baryons $B^{ij}$, $\tl B_{\tli\tlj}$
with the Hodge-duals of magnetic baryons $\eps^{ijkl}C_{kl}$ and
$\eps_{\tli \tlj \tl k \tl l}\tl C^{\tl k \tl l}$. Similarly to the
first dual, the existence of elementary mesons explicitly breaks
$SU(8)$ symmetry down to $SU(4)_L\times SU(4)_R \times U(1)_B$.

The third dual theory was proposed by Intriligator and Pouliot
in~\cite{IP}, where they generalise the original Seiberg series of
dual theories to $SP(N_c)$ gauge groups. The full set of electric
colour singlets is added as fundamental fields, and all composite
singlets are made massive by the superpotential. Thus, the moduli
space is trivially the same as electric one, and all electric
composites are identified with corresponding elementary fields in the
dual theory. As the full set of SQCD gauge invariants fits in $SU(8)$
multiplet, this dual has $SU(8)$ global symmetry and does not exhibit
accidental symmetry in the infrared.

\section{Duality in the presence of accidental symmetry}

The first and second duals of the previous Section possess only
$SU(4)_L\times SU(4)_R\times U(1)_B$ global symmetry, which gets
promoted in the infrared to the full $SU(8)$ of the electric
theory. This is not in contradiction with the concept of Seiberg
duality which relates different theories with the same {\it infrared}
properties. The dual theories may have not only different gauge groups
and field contents, but also different global symmetries.

It is worth mentioning that the 't Hooft anomaly matching
conditions~\cite{tHooft} are somewhat peculiar in this situation. As
the flavour group of the magnetic theory is smaller than that of its
infrared descendant, anomaly matching conditions in the magnetic
theory apply to smaller set of global currents. In other words, not
all anomaly relations of the electric theory have their magnetic
counterparts, simply because the magnetic theory has smaller set of
global currents.

Once the global groups of electric theory and its magnetic dual are
different, there is an ambiguity in identifying the operators of the
two theories. Operators related by global symmetry in electric theory
may no longer have this property in magnetic theory; this ambiguity
becomes irrelevant in the infrared only.  Hence, a small deformation
of the electric theory may have several duals emanating from one and
the same undeformed magnetic master theory. As an example, as proposed
in Ref.~\cite{Leigh}, one can modify electric and magnetic theories by
introducing small terms into their superpotentials. These terms are
related by duality correspondence, and it is expected that duality
remains valid for modified theories. Above the infrared, this
correspondence is ambiguous, and unique superpotential term in theory
with larger symmetry corresponds to a family of inequivalent terms in
dual theory.

Consider small mass term for a pair of quarks in electric theory of
the previous Section.  For definiteness, take it as $m\,Q^4\cdot\tl
Q_{\tl4}$ .  Its magnetic counterpart in the first dual theory,
obtained according to the above default identification, is
$m\,q^4\cdot\tl q_{\tl4}$ (up to a constant factor, see
Eq.(\ref{dec15-09})). However, the $SU(8)$ rotation of the mass term,
innocent in electric theory, gives rise to terms $m\,B^{34}$, $m\,\tl
B_{\tl3\tl4}$ in the superpotential of the magnetic theory, as well as
their linear combinations with $m\,q^4\cdot\tl q_{\tl4}$. These cannot
be rotated back to $m\,q^4\cdot\tl q_{\tl4}$ by $SU(4)\times
SU(4)\times U(1)$ of the ``magnetic theory''.  In the infrared,
$q^4\cdot\tl q_{\tl4}$ is replaced by magnetic meson $N^4_{\tl4}$, and
all mass terms in the dual theory are related by the accidental
$SU(8)$.

Likewise, the modifications of the second dual theory in addition to
default magnetic counterpart $m\,M^4_{\tl4}$ contain also
superpotential terms $m\,q_1\cdot q_2$, $m\,\tl q^{\tl1}\cdot\tl
q^{\tl2}$ and their linear combinations with $m\,M^4_{\tl4}$. These
cannot be related to each other by $SU(4)\times SU(4)\times U(1)$
transformations.

We see that once introduced, the mass term leads to a family of
inequivalent dual theories emanating from one master theory.  In this
sense, $N_c = 2$, $N_f = 4$ SQCD \emph{with small mass term} for quark
flavour has two continuous families of dual descriptions, namely, the
first and second magnetic duals with inequivalent families of terms in
the superpotentials.

Our point is that it is this kind of multiplicity that has been found
by SV using superconformal index matching. For $N_c = 2$, $N_f = 4$
SQCD, the matching approach suggests that there are 72 theories dual
to each other. These include the original electric theory, the third
dual from our table and two sets of 35 dual theories corresponding to
the first and second magnetic duals. Duals in each set have identical
field content and Lagrangian, namely, those of the first or second
magnetic theories. The only difference between them is the way the
anomalies match with electric theory, i.e., the way $SU(4)_L\times
SU(4)_R\times U(1)_B$ is embedded into $SU(8)$.  This part of
multiplicity is precisely due to the ambiguity in relating operators
in dual theories.

As they stand, all 35 duals are equivalent to their master theory.  To
make this multiplicity physical, one deforms the electric theory.
Instead of using $SU(8)$ breaking terms in superpotential as proposed
in~\cite{Leigh}, we introduce external gauge field that weakly couples
to global current of the Cartan subgroup of $SU(8)$.  Then the family
of dualities found by SV in the superconformal index context are
dualities between the theories \emph{in the external field} coupled to
inequivalent global currents.

To proceed further, we note that $SU(4)_L\times SU(4)_R\times U(1)_B$
of the first and second magnetic theories is embedded in electric
$SU(8)$ in non-trivial way.

Consider $SU(8)$ Cartan generator represented on electric quarks by
$diag(x_i,\tl x_j)$ with $i,j = 1,2,3,4$. Let us find its
representation on magnetic quarks using the identification of gauge
invariants. Suppose that this operator is represented on magnetic
quarks and elementary baryons of the first dual theory by
$diag(z_i,\tl z_j)$ and $diag(y_i,\tl y_j)$, respectively. Demanding
that the elementary magnetic baryons have the same charges as
composite electric ones, we obtain $x_i = y_i$ and $\tl x_j = \tl
y_j$. Invariance of the superpotential terms $\eps_{ijkl}B^{ij}\,q^k
\cdot q^l$ and $\eps^{\tli \tlj \tl k \tl l} \tl B_{\tli \tlj}\,\tl
q_{\tl k} \cdot \tl q_{\tl l}$ (see Table~\ref{Table}) gives for the
charges of the magnetic quarks
\begin{align*}
z_i  & = y_i - \half \sum_k y_k = x_i - \half \sum_k x_k\,,\\
\tl z_j  & = \tl y_j - \half \sum_l \tl y_l = \tl x_j - \half \sum_l
\tl x_l\,.
\end{align*}
Charges of magnetic quarks of the second dual theory
with elementary mesons are obtained in a similar way,
\begin{align*}
z_i  & = - x_i + \half \sum_k x_k\,,\\
\tl z_j  & = -\tl x_j + \half \sum_l\tl x_l\,.
\end{align*}
Remarkably, these relationships were found by SV as transformations
acting on arguments of superconformal indices that lead to the duality
identities. The arguments of index are precisely the global group
elements, parametrized by eigenvalues of their matrix
representation. It turns out that these transformations together with
permutations of quarks generate the Weyl group $W(E_7)$ of the
exceptional root system $E_7$, which defines transformational
properties of hyperelliptic integrals that give indices of dual
theories.

Consider now baryon charge represented on electric quarks  by 
\[
Q_B = diag(+1,+1,+1,+1,-1,-1,-1,-1) \ .
\]
With the above identifications, we
see that magnetic quarks of the first (second) dual theory
have the opposite (same) baryon charges as electric ones, in
accord with Table~\ref{Table}. There are two other representations of
baryon charge on electric quarks that lead to inequivalent
representations on magnetic quarks. We choose them in electric
theory
as follows,
\begin{align*}
Q_B' & = diag(+1,+1,+1,-1,-1,-1,-1,+1)\ ,\\
Q_B''&  = diag(+1,+1,-1,-1,-1,-1,+1,+1)\ .
\end{align*}
The corresponding representations on magnetic quarks are
\begin{align*}
Q'_{B1} & = diag(0,0,0,-2,0,0,0,+2)\ ,\\
Q''_{B1} & = diag(+1,+1,-1,-1,-1,-1,+1,+1)
\end{align*}
for the first dual and
\begin{align*}
Q'_{B2} & = diag(0,0,0,+2,0,0,0,-2)\ ,\\
Q''_{B2} & = diag(-1,-1,+1,+1,+1,+1,-1,-1)
\end{align*}
for the second dual.

Coupling of external field to the baryon current breaks $SU(8)$
symmetry of electric theory down to $SU(4)\times SU(4)\times
U(1)_B$. The pattern of symmetry breaking in magnetic theories depends
on the representation of baryon charge.  Coupling of the external
field to the charge $Q_B$ does not break $SU(4)\times SU(4)\times
U(1)_B$ symmetry at all. Coupling to $Q'_B$ and $Q''_B$ breaks each of
$SU(4)$ down to $SU(3)\times U(1)$ and $SU(2)\times SU(2)\times U(1)$,
respectively. Hence, the resulting magnetic theories in external field
are physically different.  The fact that anomalies of weird baryon
charges $Q'_B$ and $Q''_B$ match to those of electric baryon current
gives extra evidence that magnetic theory possesses the full $SU(8)$
symmetry in the infrared.

Thus, upon coupling different representations of the baryon current to
external field, each of the two duals split into three inequivalent
theories.  These are the Seiberg duals to electric SQCD in external
field coupled to the baryon current. Note that one of the above
dualities is precisely the explicit example given by SV in Tables 5, 6
of~\cite{VSp:oct}; the fields are split into representations of
\[
SU(3)_L\times U(1)_L\times SU(3)_R\times U(1)_R\times U(1)_B
\]
and $U(1)_B$ charges coincide with our $Q'_B$. Electric
  counterparts of the $U(1)_{L/R}$ charges are given by
\begin{align*}
Q_L &= diag(-1,-1,-1,0,0,0,0,3)\\
Q_R &= diag(0,0,0,3,-1,-1,-1,0)\ .
\end{align*}
In magnetic theories, these are precisely the $U(1)$-subgroups
remaining after $SU(4)_{L/R}$ split into $SU(3)\times U(1)$ in
external field interacting with the charge $Q'_B$.

Considering in this way all other Cartan generators\footnote{In fact,
it is sufficient to consider the least symmetric Cartan generators
that break $SU(8)$ down to $[U(1)]^7$.}, coupled to external fields it
is straightforward to obtain the two complete sets of 35 dual theories
suggested by SV.  They correspond to $\half C^4_8$ splittings of eight
eigenvalues into two groups acting on left and right magnetic quarks,
respectively, modulo interchanging the notions of left and
right. Remarkably, there is one-to-one correspondence between this
procedure and the index approach: no extra dualities are obtained in
our field-theoretic way as compared to SV.

We see that multiplicity of dual descriptions suggested by
superconformal index matching reveals some new aspects of duality. We
have found that most of 72 dual theories suggested by superconformal
index matching can be considered independent only in quite
unconventional sense, namely, by coupling different global currents of
the same theory to external field. Without this additional
construction, there are only four theories dual to each other. It is
worth noting, though, that several aspects of multiple duality, like
the r\^ole of $W(E_7)$ transformations that somehow relate all four
duals, remain unclear.

\section{Reduction to $N_f = 3$}

One way of studying known dualities and obtaining new ones is to
consider the flow of dual theories under integration out some of the
fields. SV proposed a reduction procedure for elliptic hypergeometric
integral identities that starts with an established relation between
superconformal indices of two theories and gives a relation for
theories with lower rank of global symmetry group. One expects that
this reduction corresponds to integration out some of the matter
fields in pertinent field theories. SV considered the reduction that
effectively removes one quark flavour from $N_c = 2$, $N_f = 4$ SQCD
and its duals. As a result, a set of 36 gauge theories dual to each
other was proposed. Only tree of them have different field
content. One of these theories is $N_f = 3$, $N_c=2$ SQCD without
extra fields, while two others are $N_f = 3$, $N_c=2$ SQCD with two
different sets of additional gauge singlets.

At the same time, it is known that electric and the second and third
dual theories flow to the same theory in the infrared. The low energy
theory does not have gauge symmetry and is a theory of one $SU(6)$
antisymmetric tensor field $V$ with superpotential proportional to
Pfaffian of $V$~\cite{Seiberg:exact}.

These two views on the result of integrating out one quark flavor from
$N_c = 2$, $N_f = 4$ SQCD are in apparent contradiction.  To see what
happens, let us systematically study how the theories listed in
Table~\ref{Table} flow towards the infrared. Of particular interest
are the theories that emerge as intermediate steps: we will see in
Section~\ref{sub2} that some of them are precisely the duals suggested
by the index matching approach.

\subsection{Flowing to the infrared}

Let us leave aside for the time being the conjectures based on
superconformal index approach, and integrate out one quark flavor and
its magnetic counterparts in all four theories described in Section 2.
For the second and the third dual descriptions this procedure was
briefly described in the original papers~\cite{Seiberg, IP}. Here we
follow the work~\cite{Leigh} where identifications of operators in
electric and magnetic theories were discussed.

We are going to add either the mass term for the fourth flavor,
$m\,Q^4\cdot\tl Q_{\tl4}$, or oblique mass term $h\,Q^3\cdot
Q^4$. Although these terms are related by $SU(8)$ transformation in
electric theory, they have different counterparts in the first dual
theory with the default identification of the operators. This has been
discussed in Section~3. Accordingly, the flow of the first dual theory
to the infrared proceeds through different steps for the two mass
terms. The same applies to the second dual theory. One can see that
our choice of the mass terms exhausts all possible patterns of flow;
considering other mass terms adds nothing new.

We discuss electric theory first.  Keeping explicit $SU(4)_L\times
SU(4)_R\times U(1)_B$ part of $SU(8)$ global symmetry, we begin with
the mass term $m\,Q^4\cdot\tl Q_{\tl4}$. For $m\gg\Lambda$, the
dynamical scale of electric theory, massive quarks are integrated out,
and one arrives at SQCD with $N_f = 3$ flavours. Below its own
dynamical scale $\Lambda_3 = (m\,\Lambda^2)^{1/3}$ this theory
confines and generates dynamical superpotential~\cite{Seiberg:exact}
\be\label{Wdyn}
W_{dyn} = \frac1{m\Lambda^2} \left\{
\frac14\eps_{rst}\, B^{rs} M^t_{\tl t}\,\tl B_{\tl r \tl s}\,\eps^{\tl
r \tl s \tl t} - \det M \right\} = \frac{\text{Pf } V}{\Lambda_3^3}\ .
\ee
Here $r,\tl r, \ldots = 1,2,3$ are the indices of $SU(3)_L\times
SU(3)_R$ flavour group that remains after integration out. The actual
full global symmetry is $SU(6)$, and $V$ is antisymmetric $SU(6)$
tensor composed of mesons and baryons.

Let us now consider oblique mass term $h\,Q^3\cdot Q^4$. It breaks
flavour group in another manner: $SU(4)_L\times SU(4)_R\times U(1)_B
\subset SU(8) \to SU(2)_L\times SU(4)_R\times U(1) \subset
SU(6)$. Integrating out massive quarks, we arrive at SQCD with two
surviving left quarks $Q^a$, $a=1,2$ and four right quarks $\tl Q_{\tl
j}$. This is still $N_f = 3$ SQCD.  Below the scale
$(h\,\Lambda^2)^{1/3}$ quarks are confined in mesons $M^a_{\tl j}$,
baryons $B^{ab}$ and antibaryons $\tl B_{\tli\tlj}$. Dynamically
generated superpotential is given by Eq.~\eqref{Wdyn} with $\tl
Q_{\tl4}$ substituted for $Q^3$:
\be
\label{WdynB} W_{dyn} =
\frac1{4h\Lambda^2} \left\{M^a_{\tli} M^b_{\tlj} - \frac14 B^{ab} \tl
B_{\tli\tlj}\right\} \eps^{\tli \tlj \tl k \tl l}\, \eps_{ab}\, \tl
B_{\tl k \tl l}\ .
\ee
This description also possesses the full
$SU(6)$ global symmetry. It is equivalent to the previous one after
field redefinition, as should be the case since the two low energy
descriptions originate from one and the same theory with mass terms
related by $SU(8)$ transformation.

Thus, the electric theory flows through $N_f = 3$ SQCD with no gauge
singlet fields, whose low energy description is the theory of $SU(6)$
antisymmetric tensor field.

\vskip2em
%First dual
%%usual mass

Let us consider what happens with the first dual theory as we add each
of the mass terms. Before writing counterparts of the mass terms we
note one subtlety in the operator identification. While electric
composite baryons and antibaryons match the corresponding elementary
magnetic fields directly\footnote{Because of different canonical
dimensions of these fields, this matching involves a scale $\mu$.}
(see Table~\ref{Table}), composite mesons match up to non-trivial
factor:
\[
M^i_{\tlj} \simeq \sqrt{\frac{-\Lambda^2}{\mu^2}}\,N^i_{\tlj} 
\simeq \sqrt{\frac{-\Lambda^2}{\mu^2}}\, q^i\cdot\tl q_{\tlj}\ ,
\]
where the scale $\mu$ is introduced on dimensional grounds.  This
identification ensures that the duality transformation applied twice
gives back the original theory.

With this qualification, the electric theory with mass term
$m\,Q^4\cdot\tl Q_{\tl4}$ is dual to the the magnetic one with the
following superpotential,
\be
W = \frac1{4\mu}\,\eps_{ijkl}B^{ij}\,q^k \cdot q^l +
\frac1{4\mu}\,\eps^{\tli \tlj \tl k \tl l} \tl B_{\tli \tlj}\, \tl
q_{\tl k} \cdot \tl q_{\tl l} +
m\sqrt{\frac{-\Lambda^2}{\mu^2}}\,q^4\cdot\tl q_{\tl4}\ .
\label{dec15-09}
\ee
For large $m$, quarks $q^4$ and $\tl q_{\tl4}$ are heavy. Integrating
them out we arrive at the theory with $N_f =3$ quark flavours and the
same set of elementary baryons as in the original magnetic $N_f = 4$
theory. The tree level superpotential after integration out is given
by
\be\label{Wtree}
W_{tree} = \frac1{2\mu}\,\eps_{rst}\,q^r\cdot q^s B^{t4}+
\frac1{2\mu}\,\eps^{\tl r\tl s\tl t} \tl q_{\tl r}\cdot\tl q_{\tl s}\,
\tl B_{\tl t \tl4} - \frac1{4m\mu^2}\sqrt{\frac{\mu^2}{-\Lambda^2}}\,
\eps_{rst}\,\eps^{\tl r\tl s\tl t} B^{rs} \tl B_{\tl r\tl s}\,q^t\cdot
\tl q_{\tl t}\ ,
\ee
 where all indices take the values $1,2,3$. This theory has its own
scale given by
\[
\tl\Lambda_3 \equiv \l m\,\frac{\Lambda}{\mu}\,\tl \Lambda^2\r^{1/3} =
\l m\,\tl\Lambda\,\mu\r^{1/3}\ ,
\]
where $\tl \Lambda \equiv \mu^2/\Lambda$ is the scale of the initial
magnetic theory. Below this scale the theory confines and generates
dynamical superpotential analogous to~\eqref{Wdyn}:
\[
W_{dyn} = \frac1{m\tl \Lambda^2}\,\sqrt{\frac{\mu^2}{-\Lambda^2}}\,
\left\{ \frac14\,\eps_{rst}\,C^{rs}\,N^t_{\tl t}\, \tl C_{\tl r \tl
s}\,\eps^{\tl r \tl s \tl t} - \det N \right\}\ ,
\]
where $C^{rs} \equiv q^r\cdot q^s$ and $\tl C_{\tl r \tl s} \equiv \tl
q_{\tl r}\cdot\tl q_{\tl s}$ are composite magnetic baryons and
antibaryons of $N_f = 3$ theory.

Upon rescaling the meson field $N^i_{\tlj}$ to match the electric one
$M^i_{\tlj}$, we obtain the theory of interacting mesons and baryons
(the latter are $B^{rs}$, $B^{r4}$, $C^{rs}$ and their antibaryons)
with the superpotential given by the sum of $W_{dyn}$ and $W_{tree}$:
\begin{multline}\label{WCsaki}
W =  \frac1{m\Lambda^2} \left\{ \frac14\eps_{rst}\,  B^{rs}
M^t_{\tl t}\,\tl B_{\tl r \tl s}\,\eps^{\tl r \tl s \tl t} - \det M
\right\} + \\
\frac1{2\mu}\,\eps_{rst}\,C^{rs} B^{t4}+
\frac1{2\mu}\,\eps^{\tl r\tl s\tl t}\,\tl C_{\tl r\tl s}\, \tl
B_{\tl t \tl4} - \frac1{4m\mu^2}\,\eps_{rst}\,C^{rs}\,M^t_{\tl t}\,\tl
C_{\tl r \tl s}\, \eps^{\tl r \tl s \tl t}\ . 
\end{multline}
According to our definition of $B$, the canonically normalized field
is $B/\mu$. Since $C$ is the composite field arising because of strong
coupling at $\tl\Lambda_3$, its canonically normalized form is
$C/\tl\Lambda_3 \simeq q\cdot q/\tl\Lambda_3$. Hence, this
superpotential gives masses of the same scale $\tl \Lambda_3$ to all
baryons absent in electric $N_f = 3$ theory. Integrating them out we
are left with the same theory as in the electric case with the
superpotential given by the first line of Eq.~\eqref{WCsaki},
cf. Eq.~\eqref{Wdyn}.

Hence, the electric theory and its first dual flow to the same theory
of interacting singlets. Notably, the flow of the first dual proceeds
through the intermediate description. This is $N_f = 3$ SQCD with six
baryons $B^{rs}$, $B^{r4}$, six antibaryons $\tl B_{\tl r\tl s}$, $\tl
B_{\tl s\tl4}$ and superpotential given by Eq.~\eqref{Wtree}. The
flavour group of this theory is
\be\label{addi-1}
SU(3)_L\times SU(3)_R\times U(1)_B\times U(1)_4 \ ,
\ee
where the last factor is inherited from the initial magnetic theory
together with extra baryons  $B^{r4}$, $\tl B_{\tl
s\tl4}$. In electric
theory, this symmetry acts only on massive quarks and disappears once
these quarks have been integrated out.

\vskip1em
%oblique mass

Let us now turn to the second electric mass term $h\,Q^3\cdot Q^4$.
It  corresponds to the
term $h\,B^{34}$ in the first magnetic dual. The full
superpotential is now given by
\be
\label{Wh}
W = \frac1{4\mu}\,\eps_{ijkl}B^{ij}\,q^k \cdot q^l + h\,B^{34}+
\frac1{4\mu}\,\eps^{\tli \tlj \tl k \tl l} \tl B_{\tli \tlj}\, \tl
q_{\tl k} \cdot \tl q_{\tl l}\ .
\ee
It generates non-zero expectation value for magnetic quarks,
\be
\label{vev} 
q^1\cdot q^2 = -\mu h \equiv v^2\ , 
\ee 
which
completely breaks the gauge group. Let us choose the quark vev to be
$(q^a)_{\al} = v\,\delta^a_{\al}$, where $\al,\ldots = 1,2$ is
colour index. Then for $h \gg \tl \Lambda$, the fields $(q^a)_a$ as well as
$(q^3)_{\al}$, $(q^4)_{\al}$ and baryons, except for $B^{12}$,
obtain masses and are integrated out. The fields $(q^1)_2$,
$(q^2)_1$ and $B^{12}$ remain massless but completely decouple. Thus
we are left with eight quark components $(\tl q_{\tlj})_\al$
and antibaryons $\tl B_{\tli \tlj}$ with superpotential given by the
last term in Eq.~\eqref{Wh}.

Using the quark vev we rescale quark fields to match them to
electric mesons:
$$
(\tl q_{\tli})_\al \equiv
-\eps_{\al b}\,\frac1v\,\sqrt{\frac{\mu^2}{-\Lambda^2}}\,M^b_{\tli}\ .
$$
Tree level superpotential now reads
$$
W_{tree} =  \frac1{4h\Lambda^2}\,\eps_{ab}\,
M^a_{\tli}\,M^b_{\tlj}\, 
\tl B_{\tl k \tl l}\,\eps^{\tli\tlj \tl k \tl l}\ .
$$
It matches  the first term in Eq.~\eqref{WdynB}.

The remaining part of superpotential is generated dynamically. To check
this, one  gives  large vevs to all baryon and antibaryon fields
except for $B^{34}$. Due to superpotential~\eqref{Wh}, baryon vevs provide
masses to all quarks except for $q^1$ and $q^2$. Integrating the massive 
quarks
out one obtains a theory with one quark flavour composed of $q^1$
and $q^2$ and unbroken $SU(2)$ colour group. This theory confines and
generates non-perturbative dynamical superpotential~\cite{Seiberg:exact}
$$
W_{dyn} = \frac{(\tl \Lambda_1)^5}{\langle q^1\cdot q^2\rangle}.
$$
Here $\tl \Lambda_1$ is the scale of the resulting $N_f = 1$ theory. It
is determined by baryon vevs and the scale of original theory,
$$
(\tl \Lambda_1)^5 = \text{Pf }\left[B^{ab}\oplus\tl B_{\tli\tlj}\right]
\mu^{-3} \tl \Lambda^2 =
\frac{\mu}{16\Lambda^2}\,\eps_{ab}\,B^{ab}\,\eps^{\tli \tlj \tl k \tl
  l}\,\tl B_{\tli\tlj}\,\tl B_{\tl k \tl l}\ ,
$$
where we have used the relation $\Lambda\tl\Lambda = \mu^2$ between the
scales of electric and magnetic theories.
Using this expression and the quark vev~\eqref{vev}, one obtains the
dynamically generated superpotential
\[
W_{dyn} = - \frac1{16h\Lambda^2}\, \eps_{ab}\, B^{ab} \eps^{\tli \tlj
\tl k \tl l}\,\tl B_{\tli\tlj}\,\tl B_{\tl k \tl l}\ ,
\]
which is precisely the second term in Eq.~\eqref{WdynB}. In this
way we arrive at the same theory of the antisymmetric $SU(6)$ tensor
field we have obtained starting from electric theory.  As in the
previous case, part of the superpotential is given by tree-level terms
while the other part is generated non-perturbatively. As opposed to
the previous case, however, the term $h\,B^{34}$ breaks the gauge
group and the theory does not pass through any intermediate
description during the flow.

\vskip2em
%second dual
%%usual mass

Now let us turn to the second dual theory. Adding the electric mass
term $m\,Q^4\cdot\tl Q_{\tl4}$ induces the term $m\,M^4_{\tl4}$ in
this magnetic dual. The resulting flow resembles closely that of the
first dual with the term $h\,B^{34}$.  As described by
Seiberg~\cite{Seiberg}, the term $m\,M^4_{\tl4}$ in superpotential
leads to non-vanishing quark vev $\langle q_4\cdot \tl
q^{\tl4}\rangle$, which completely breaks the gauge symmetry. The
quark components that remain massless are identified with electric
baryons. The superpotential arises here as the sum of dynamical and
tree-level contributions, just as in the case considered above. This
flow does not proceed through intermediate descriptions.

\vskip1em
%%oblique mass term

The oblique mass term $h\,Q^3\cdot Q^4$ in electric theory leads to
more interesting flow of the second dual.  It was considered in detail
in Ref.~\cite{Leigh}. This flow is similar to that of the first dual
under addition of the mass term
$m\,\sqrt{\frac{-\Lambda^2}{\mu^2}}\,q^4\cdot\tl q_{\tl4}$.  In the
second dual, the electric baryons match to the magnetic ones as
\be\label{bmatch}
B^{ij} \simeq \sqrt{\frac{-\Lambda^2}{\mu^2}}\,\eps^{ijkl}\,C_{kl}
\equiv \sqrt{\frac{-\Lambda^2}{\mu^2}}\,\eps^{ijkl}\,q_k\cdot q_l \ ,
\ee
and the same for antibaryons. Thus, electric mass term $h\,Q^3\cdot
Q^4$ corresponds to the term 
$$
h\,\sqrt{\frac{-\Lambda^2}{\mu^2}}\,q_1\cdot q_2 \ .
$$
For $h \gg \tl\Lambda$, it gives  mass to quarks $q_1$ and
$q_2$. This term breaks $SU(4)_L$ part of the flavour
symmetry to diagonally embedded $SU(2)_{12}\times SU(2)_{34}$.  The full
superpotential now reads
$$
W = \frac1{\mu}\,M^i_{\tli}\,q_i\cdot\tl q^{\tlj} +
h\,\sqrt{\frac{-\Lambda^2}{\mu^2}}\,q_1\cdot q_2\ .
$$
Integrating out massive quarks one obtains SQCD with $N_f = 3$
quark flavours, mesons $M^i_{\tlj}$ and tree-level superpotential
\be
W_{tree} = \frac1{\mu}\,M^f_{\tli}\,q_f\cdot\tl q^{\tlj} -
\frac1{2h\mu^2}\,\sqrt{\frac{\mu^2}{-\Lambda^2}}\,\eps_{ab}\,
M^a_{\tli}M^b_{\tlj}\,\tl q^{\tli}\cdot\tl q^{\tlj}\ ,
\label{addi-3}
\ee
where $a,b = 1,2$ and $f,g = 3,4$ are the indices of the
representations of $SU(2)_{12}$ and $SU(2)_{34}$, respectively. At
this stage the theory has global symmetry
\be
\label{addi-2}
SU(2)_{12}\times SU(2)_{34}\times SU(4)_R\times U(1)_B
\ee
where $U(1)_B$ is a combination of the initial baryon charge and 
the third Cartan generator of $SU(4)_L$ that does not belong to
 $SU(2)_{12}$ or $SU(2)_{34}$. In comparison to the electric $N_f = 3$
SQCD, this theory has extra $SU(2)_{12}$ symmetry. This symmetry is
the magnetic counterpart of electric $SU(2)$ 
that acts on  quarks
$Q^3$ and $Q^4$ and is preserved by the oblique mass term $h\,Q^3\cdot Q^4$. 
This electric $SU(2)$
completely disappears after the heavy quarks are integrated out. As
opposed to electric theory, the intermediate magnetic theory
contains mesons that transform non-trivially under $SU(2)_{12}$.
Hence,  $SU(2)_{12}$ is
 a non-trivial part of global symmetry of the second dual theory
at this stage of its flow.

SQCD sector of this theory has $N_f = N_c + 1$ flavors,
and thus confines. The
additional dynamical superpotential for its own composite
mesons $N^{\tli}_f$ and baryons $C_{fg}$, $\tl C^{\tli\tlj}$ is generated.
The baryon fields  $C_{fg}$, $\tl C^{\tli\tlj}$ can be replaced 
by
$B^{ab}$, $\tl B_{\tli\tlj}$ by making use of the
field redefinition~\eqref{bmatch}.  In this way we obtain the
theory of baryons $B^{ab}$, antibaryons $\tl B_{\tli\tlj}$ and 
mesons $M^a_{\tli}$, $M^f_{\tli}$, $N^{\tli}_f$ with 
superpotential 
\begin{multline}\label{WSeiberg}
W_{dyn} = \frac1{4h\Lambda^2} \left\{M^a_{\tli} M^b_{\tlj} - \frac14
B^{ab} \tl B_{\tli\tlj}\right\} \eps^{\tli \tlj \tl k \tl l}\,
\eps_{ab}\, \tl B_{\tl k \tl l} +\\
\frac1{\mu}\,M^f_{\tli}\,N^{\tli}_f +
\frac1{2h\mu^2}\,\eps^{fg}\,N^{\tli}_f\,N^{\tlj}_g\,\tl B_{\tli\tlj} \ . 
\end{multline}
This superpotential gives masses to all fields that are absent in the
electric low energy description, namely, to $M^f_{\tli}$ and
$N^{\tli}_{f}$.  After integrating them out, the remaining fields fit
into $SU(6)$ antisymmetric tensor representation, and the theory is
equivalent to the common low energy description we obtained in
previous cases.

\vskip2em
%third dual

For completeness, let us describe the flow of the third dual
description~\cite{IP}. It possesses full $SU(8)$ flavour
symmetry. Thus, any electric mass term has the same effect in the dual
theory.  It is easier to consider the flow of this theory in $SU(8)$
notations. The matter fields are magnetic quarks $q_I$, $I =
1,\ldots8$, and elementary gauge singlet antisymmetric tensor $V^{IJ}$
that contains all mesons and baryons. The latter corresponds to the
moduli of electric theory.  Superpotential of this dual takes the form
\[
W = \frac1\mu\,V^{IJ}\, q_I\cdot q_J\ .
\] 
Modulo
$SU(8)$ transformation, the
magnetic counterpart of any  electric mass term is $m\,V^{12}$. 
With
this term in superpotential,  quarks $q_1$ and $q_2$ obtain vev
\mbox{$\langle q_1\cdot q_2 \rangle = - \mu\,m$} that breaks
the gauge group. Then all quarks 
and the fields
$V^{1I}$ and
$V^{2I}$ become massive.
 Integrating them out one is left with
the remaining components of $V^{IJ}$ which transform as antisymmetric tensor
under the surviving $SU(6)$ flavour group, and no tree-level
superpotential. Superpotential~\eqref{Wdyn} is  generated by
instantons of the broken gauge group as described in~\cite{Seiberg:exact}.

%SV matching and conclusion

Hence, once mass term of any form is added, every theory given in
Table~\ref{Table} flows to one and the same theory of $SU(6)$
antisymmetric tensor field with superpotential given by
Eq.~\eqref{Wdyn}.  This flow, however, is different for different
duals and for different forms of the mass term.

\subsection{Making contact with index matching}
\label{sub2}

Finally, let us make contact with dualities for $N_f = 3$ SQCD
suggested by SV using the reduction of superconformal index identities
of $N_f = 4$.  The conjecture is that there exist 36 dual gauge
theories, apart from the low energy description in terms of the
antisymmetric $SU(6)$ tensor. All of them are $SU(2)$ SQCD with $N_f =
3$ flavours of quarks and additional singlet fields. Apart from the
obvious electric $N_f = 3$ SQCD, there are two options for the field
content and global symmetry. All other theories are different only in
the sense explained in the previous Section.

In notations of Ref.\cite{VSp:oct}, one of these two conjectured duals
possesses $SU(3)_L\times SU(3)_R\times U(1)_B\times U(1)_{add}$ global
symmetry and besides SQCD sector contains 12 gauge singlet fields:
$M_1$ in $\mathbf{3_A}$ and $N_1$ in $\mathbf{3}$ of $SU(3)_L$, and
$M_2$ in $\mathbf{3_A}$ and $N_2$ in $\mathbf{3}$ of $SU(3)_R$, see
Table 12 of~\cite{VSp:oct}. This theory is nothing but the theory
obtained as an intermediate step in the flow of the first dual. In our
notations, $M_1$, $N_1$ are baryons $B^{rs}$, $B^{r4}$, and $M_2$,
$N_2$ are antibaryons $\tl B_{\tl r\tl s}$, $\tl B_{\tl r\tl4}$. The
group $U(1)_{add}$ is $U(1)_4$ entering (\ref{addi-1}). In addition,
we have found that the superpotential of this theory is given
by Eq.~\eqref{Wtree}.

Like its progenitor, the first dual for $N_f = 4$ SQCD, this theory
has lower flavour group $SU(3)\times SU(3)\times U(1)_B\times U(1)_4$
in comparison with $SU(6)$ of the electric $N_f = 3$ SQCD. Thus, in
the same way as in the previous Section, multiple dualities emerge in
this theory when global currents are coupled to external field.  In
this case, there are 20 duals that correspond to $C^3_6$ ways to split
six eigenvalues of the Cartan generator of the $SU(6)$ global symmetry
into those acting on left and right magnetic quarks.  We note here
that unlike in $N_f=4$ case, interchanging left and right quarks in
the magnetic theory does not leave that theory intact, as it changes
the sign of the $U(1)_4$ charge of baryons $B^{r4}$; hence the above
counting.

The second dual proposed by SV possesses $SU(2)_L\times
SU(2)_{add}\times SU(4)_R\times U(1)_B$ global symmetry with 16 gauge
singlet fields: $M$ in $(\mathbf{1, 2, 4})$ representation and $N$ in
$(\mathbf{2,1, 4})$ representation (cf. Table 13
of~\cite{VSp:oct}). We identify this theory with the intermediate
description obtained during the flow of the second $N_f = 4$ dual
theory with $M$ identified with the mesons $M^f_{\tli}$ and $N$ with
the mesons $M^a_{\tli}$, see (\ref{addi-2}).  Its tree-level
superpotential is given by (\ref{addi-3}).

By coupling global currents of this theory to external field one
obtains 15 dualities, in the sense of the previous Section. These
correspond to $C^2_6$ ways to label six eigenvalues of the Cartan
generator of $SU(6)$ as two left and four right.

As a result, the total number of dualities (in external field sense)
is $1 + 20 + 15 = 36$, taking into account electric $N_f = 3$
SQCD. Without the external field, there are two new dual descriptions
for $N_c=2$, $N_f = 3$ SQCD conjectured from index matching and
supported by our study.  Both are valid Seiberg dual theories, which
show identical infrared behaviour. These descend from $N_c=2$, $N_f =
4$ SQCD dualities by integrating out one flavour, and are $N_c=2$,
$N_f=3$ SQCD with additional singlets. Tree level
superpotentials in the two theories are given by (\ref{Wtree}) and
(\ref{addi-3}), respectively. Unlike in many other cases, one can
explicitly follow the flow of all three theories to the infrared and
check that they have one and the same low energy description.

\section{Acknowledgments}

The author is indebted to S.~Demidov, D.~Gorbunov, V.~Spiridonov,
V.~Vartanov and especially to V.~Rubakov for useful discussions. The
work has been supported in part by the Russian Foundation for Basic
Research grant 08-02-00473a, by the grants of the President of the
Russian Federation MK-4317.2009.2, NS-1616.2008.2 (government contract
02.740.11.0244), by FAE program (government contract P520), by the
`MassTeV' ERC advanced grant 226371, and by the `Dynasty' foundation.

\end{document}